\PassOptionsToPackage{unicode}{hyperref}
\PassOptionsToPackage{hyphens}{url}
\PassOptionsToPackage{dvipsnames,svgnames,x11names}{xcolor}
\documentclass[
  10pt,
  letterpaper,
  twocolumn]{article}
\usepackage{xcolor}
\usepackage{amsmath,amssymb}
\usepackage{iftex}
\ifPDFTeX
  \usepackage[T1]{fontenc}
  \usepackage[utf8]{inputenc}
  \usepackage{textcomp} 
\else 
  \usepackage{unicode-math} 
  \defaultfontfeatures{Scale=MatchLowercase}
  \defaultfontfeatures[\rmfamily]{Ligatures=TeX,Scale=1}
\fi
\usepackage{lmodern}
\ifPDFTeX\else
\fi
\IfFileExists{upquote.sty}{\usepackage{upquote}}{}
\IfFileExists{microtype.sty}{
  \usepackage[]{microtype}
  \UseMicrotypeSet[protrusion]{basicmath} 
}{}
\makeatletter
\@ifundefined{KOMAClassName}{
  \IfFileExists{parskip.sty}{%
    \usepackage{parskip}
  }{
    \setlength{\parindent}{0pt}
    \setlength{\parskip}{6pt plus 2pt minus 1pt}}
}{
  \KOMAoptions{parskip=half}}
\makeatother
\makeatletter
\ifx\paragraph\undefined\else
  \let\oldparagraph\paragraph
  \renewcommand{\paragraph}{
    \@ifstar
      \xxxParagraphStar
      \xxxParagraphNoStar
  }
  \newcommand{\xxxParagraphStar}[1]{\oldparagraph*{#1}\mbox{}}
  \newcommand{\xxxParagraphNoStar}[1]{\oldparagraph{#1}\mbox{}}
\fi
\ifx\subparagraph\undefined\else
  \let\oldsubparagraph\subparagraph
  \renewcommand{\subparagraph}{
    \@ifstar
      \xxxSubParagraphStar
      \xxxSubParagraphNoStar
  }
  \newcommand{\xxxSubParagraphStar}[1]{\oldsubparagraph*{#1}\mbox{}}
  \newcommand{\xxxSubParagraphNoStar}[1]{\oldsubparagraph{#1}\mbox{}}
\fi
\makeatother

\usepackage{longtable,booktabs,array}
\usepackage{calc} 
\usepackage{etoolbox}
\makeatletter
\patchcmd\longtable{\par}{\if@noskipsec\mbox{}\fi\par}{}{}
\makeatother
\IfFileExists{footnotehyper.sty}{\usepackage{footnotehyper}}{\usepackage{footnote}}
\makesavenoteenv{longtable}
\usepackage{graphicx}
\makeatletter
\newsavebox\pandoc@box
\newcommand*\pandocbounded[1]{
  \sbox\pandoc@box{#1}%
  \Gscale@div\@tempa{\textheight}{\dimexpr\ht\pandoc@box+\dp\pandoc@box\relax}%
  \Gscale@div\@tempb{\linewidth}{\wd\pandoc@box}%
  \ifdim\@tempb\p@<\@tempa\p@\let\@tempa\@tempb\fi
  \ifdim\@tempa\p@<\p@\scalebox{\@tempa}{\usebox\pandoc@box}%
  \else\usebox{\pandoc@box}%
  \fi%
}
\def\fps@figure{htbp}
\makeatother

\NewDocumentCommand\citeproctext{}{}

\makeatletter
 \let\@cite@ofmt\@firstofone
 \def\@biblabel#1{}
 \def\@cite#1#2{{#1\if@tempswa , #2\fi}}
\makeatother
\newlength{\cslhangindent}
\newlength{\csllabelwidth}
\newenvironment{CSLReferences}[2] 
 {\begin{list}{}{%
  \setlength{\itemindent}{0pt}
  \setlength{\leftmargin}{0pt}
  \setlength{\parsep}{0pt}
  \ifodd #1
   \setlength{\leftmargin}{\cslhangindent}
   \setlength{\itemindent}{-1\cslhangindent}
  \fi
  \setlength{\itemsep}{#2\baselineskip}}}
 {\end{list}}
\usepackage{calc}

\providecommand{\tightlist}{%
  \setlength{\itemsep}{0pt}\setlength{\parskip}{0pt}}

\makeatletter
\@ifpackageloaded{caption}{}{\usepackage{caption}}
\AtBeginDocument{%
\ifdefined\contentsname
  \renewcommand*\contentsname{Table of contents}
\else
  \newcommand\contentsname{Table of contents}
\fi
\ifdefined\listfigurename
  \renewcommand*\listfigurename{List of Figures}
\else
  \newcommand\listfigurename{List of Figures}
\fi
\ifdefined\listtablename
  \renewcommand*\listtablename{List of Tables}
\else
  \newcommand\listtablename{List of Tables}
\fi
\ifdefined\figurename
  \renewcommand*\figurename{Figure}
\else
  \newcommand\figurename{Figure}
\fi
\ifdefined\tablename
  \renewcommand*\tablename{Table}
\else
  \newcommand\tablename{Table}
\fi
}
\@ifpackageloaded{float}{}{\usepackage{float}}
\floatstyle{ruled}
\@ifundefined{c@chapter}{\newfloat{codelisting}{h}{lop}}{\newfloat{codelisting}{h}{lop}[chapter]}
\floatname{codelisting}{Listing}

\makeatother
\makeatletter
\@ifpackageloaded{caption}{}{\usepackage{caption}}
\@ifpackageloaded{subcaption}{}{\usepackage{subcaption}}
\makeatother
\usepackage{bookmark}
\IfFileExists{xurl.sty}{\usepackage{xurl}}{} 
\hypersetup{
  pdftitle={LCSHBench: A Multilingual{,} Consensus-Grounded Benchmark for Library of Congress Subject Heading Assignment},
  pdfauthor={Kwok Leong Tang},
  colorlinks=true,
  linkcolor={blue},
  filecolor={Maroon},
  citecolor={Blue},
  urlcolor={Blue},
  pdfcreator={LaTeX via pandoc}}

\title{LCSHBench: A Multilingual, Consensus-Grounded Benchmark for
Library of Congress Subject Heading Assignment}
\author{Kwok Leong Tang}
\date{2026-06-01}
\begin{document}
\maketitle
\begin{abstract}
Automated subject cataloging assigns controlled-vocabulary headings to
bibliographic records, but LCSH has no standard public benchmark. We
introduce LCSHBench: 22,346 books in 15 languages from the openly
licensed Harvard, Columbia, and Princeton catalogs. Records enter only
when at least two independent cataloging agencies assigned LCSH; we
release per-catalog provenance plus union and unanimous answer views. A
concordance study of 465,187 works cataloged by all three libraries
shows why this design matters: libraries usually agree on the underlying
topic (93.3\% share a concept-level heading) but often differ in exact
expression (39.4\% have identical heading sets). LCSHBench therefore
scores both exact and concept matches, with set and rank metrics broken
down by language and heading type, across open-vocabulary generation and
full-vocabulary retrieval. As a first demonstration, a low-rank
fine-tune of a 300M on-device embedder improves cross-lingual retrieval
and beats a 3,072-dimensional hosted embedder on development exact
recall@200 (0.659 vs 0.623). The language panel shows the gain is not
uniform, and held-out-test and end-to-end confirmation remain future
work.
\end{abstract}

\section{Introduction}\label{sec-intro}

When a research library acquires a book, a professional cataloger reads
it and attaches a small set of standardised labels describing what it is
about. For English-language cataloging worldwide these labels are drawn
from the Library of Congress Subject Headings (LCSH) (Library of
Congress, n.d.), a controlled vocabulary of roughly half a million
authorised headings, many further refined by subdivisions
(e.g.~Sociology-\/-Research). Subject access of this kind is what lets a
reader find everything a collection holds on a topic, independent of the
words an author chose. Assigning it well is slow, expert work, and there
is sustained interest in whether software --- increasingly, large
language models --- can assist.

That interest is not matched by a way to measure it. Unlike the German
GND (through the LLMs4Subjects shared task (D'Souza et al. 2025)), the
biomedical MeSH vocabulary (through BioASQ ({Tsatsaronis et al.} 2015)),
or EU EuroVoc (through MultiEURLEX (Chalkidis et al. 2021)), LCSH has no
standard, public, versioned benchmark. Research groups evaluate on
private collections with private notions of correctness, so their
numbers cannot be compared, and progress cannot be tracked.

We close that gap with LCSHBench v1.0. Our contributions are:

\begin{enumerate}
\def\labelenumi{\arabic{enumi}.}
\tightlist
\item
  A consensus-grounded, multilingual dataset (22,346 books, 15
  languages) admitted on agreement among independent professional
  catalogers across three openly licensed research-library catalogs.
  Admission is record-level (\(\geq\)2 independent agencies); the default
  scoring key is the union of their headings (generous, and therefore
  including single-source headings --- about 46\% of dev assertions),
  with strict unanimous and raw per-catalog views also released so a
  study can choose its own bar.
\item
  An empirical justification for consensus. On 465,187 works cataloged
  by all three libraries, we quantify inter-cataloger agreement and show
  it has an objective concept-level core and a subjective
  expression-level surface described in Section~\ref{sec-concordance},
  directly motivating both the consensus target and the
  exact-versus-concept metric distinction.
\item
  Two precisely specified tasks --- open-vocabulary generation, and a
  retrieval pipeline evaluated over the full vocabulary rather than a
  frozen candidate pool --- with a public scorer and submission format.
\item
  A metric panel, not a headline F1 --- set and rank metrics under exact
  and concept match, sliced by language and heading type --- together
  with the empirical demonstration that the panel is necessary: the
  ranking of systems flips depending on which metric and which language
  one examines.
\item
  A demonstration that the benchmark is actionable. The panel localizes
  cross-lingual alignment as the first-stage retrieval bottleneck; a
  sub-dollar low-rank fine-tune of a 300M on-device embedder --- trained
  on records held strictly disjoint from the evaluation subset --- then
  out-retrieves a hosted commercial embedder of much larger dimension on
  first-stage development-subset recall, with the panel revealing the
  gain is specifically cross-lingual. We offer this as evidence the exam
  guides improvement, not as a final result: held-out-test and
  end-to-end evaluation are future work.
\end{enumerate}

The dataset, scorer, baselines, random seeds, and cleaning rules are
released together, so any reported result reproduces exactly.

\section{Literature review}\label{sec-related}

This work sits at the intersection of three literatures: the
cataloging-theory tradition on what a ``subject'' is and how reliably it
can be assigned; the automated subject-indexing literature on systems
that assign it; and the benchmark literature in adjacent controlled
vocabularies. We draw on all three, and locate the gap each leaves for
LCSH.

\subsection{What a ``subject'' is: subject analysis and
aboutness}\label{sec-related-aboutness}

Subject cataloging begins with subject analysis --- determining what a
work is about and rendering that judgement as controlled-vocabulary
headings (Joudrey and Taylor 2018). Hj{\o}rland's foundational work argues
that a subject is not a property simply read off a document but a
determination shaped by the analyst's purpose and frame, and develops
the related notions of aboutness, topicality, and relevance (Hj{\o}rland
1992, 2001). This matters directly for any benchmark: if ``the subject''
is partly a judgement, there is no single gold answer, and an answer key
must be built from agreement among expert judgements rather than
asserted by one. Our consensus design is a direct operationalisation of
this stance, and the concordance analysis in
Section~\ref{sec-concordance} quantifies, at scale, exactly where that
judgement is shared and where it varies.

\subsection{How much do catalogers agree? Inter-indexer
consistency}\label{sec-related-consistency}

The reliability of subject assignment has been studied for over half a
century under the heading of inter-indexer consistency. Beginning with
Zunde and Dexter's measures of consistency and quality (Zunde and Dexter
1969) and the review of two decades of studies by Leonard (Leonard
1977), this literature repeatedly finds that independent indexers agree
on the exact terms only modestly --- commonly in the 10--50\% range,
with consistency rising as terms are generalized --- including in direct
comparisons of national-library catalogers (Wolfram and Olson 2007;
Tonta 1991). Our inter-cataloger concordance in
Section~\ref{sec-concordance} is the modern, population-scale extension
of this line: across 465,187 works cataloged by three independent
libraries we recover the classic finding (exact agreement
\textasciitilde39\%) and sharpen it by separating an objective concept
core (\textasciitilde93\% concept-level agreement) from a subjective
expression surface --- the distinction that motivates both our consensus
target and our exact-versus-concept metrics.

\subsection{Automated subject indexing}\label{sec-related-automated}

Automated subject indexing has a long applied history, surveyed
comprehensively by Golub (Golub 2021). The dominant open toolkit is
Annif (Suominen 2019), which combines string-matching and supervised
extreme-multi-label learning and underpins production services at
several national libraries; recent work adapts transformer-based
extreme-multi-label classifiers to subject indexing (Bertalis et al.
2024). These systems are typically trained and evaluated against a
single institution's headings, which --- given the consistency results
above --- conflates genuine error with legitimate cataloger-to-cataloger
variation. A consensus answer key is, in part, a response to that
measurement problem.

\subsection{Large language models for subject
cataloging}\label{sec-related-llms}

The newest strand applies LLMs directly. Most relevant, Chow, Kao, and
Li (Chow et al. 2024) use ChatGPT to generate LCSH for electronic theses
and dissertations from titles and abstracts, in this paper's target
venue, concluding that LLMs can reduce cataloging time but that human
catalogers remain essential for validity, exhaustivity, and specificity.
The German National Library's DNB-AI system (Kluge and K{\"a}hler 2025)
reaches this verdict quantitatively and qualitatively at once: in
SemEval-2025 it ranked fourth by F1 but first in expert rating --- the
single most important prior result for our design, because it shows
numeric and expert rankings diverge. Chow's more recent skill-based
agentic pipeline decomposes LCSH assignment into conceptual analysis,
quantitative filtering, authority validation, and MARC field synthesis,
making explicit the domain procedure knowledge that subject-indexing
systems must operationalize (Chow 2026). Hybrid embedding-plus-LLM
pipelines are also emerging (Liu et al. 2025). Critically, each such
study uses its own private collection and its own definition of
``correct,'' so the results cannot be compared --- precisely the gap a
standard benchmark closes.

\subsection{Benchmarks in adjacent
vocabularies}\label{sec-related-benchmarks}

No LCSH-specific benchmark exists; the closest are in adjacent
vocabularies. LLMs4Subjects / SemEval-2025 Task 5 (D'Souza et al. 2025)
is the nearest template --- automated subject tagging for a German
national technical library over the GND taxonomy, with bilingual records
and a two-track (quantitative + subject-expert) evaluation, later
released as an XMTC dataset ({D'Souza et al.} 2026); we adopt its
two-track ambition and all-subjects-vs-core split and extend it from
bilingual to broadly multilingual. BioASQ ({Tsatsaronis et al.} 2015) is
the gold standard for longevity --- an annual, versioned
biomedical-indexing challenge over MeSH with a held-out evaluation
server; we borrow its versioning discipline, while noting MeSH is
smaller and more regular than LCSH and lacks free-floating subdivisions.
MultiEURLEX (Chalkidis et al. 2021) and the broader XMTC literature
establish the metric conventions we adopt --- P@k, R-Precision, and the
long-tail-label problem LCSH's \textasciitilde515K labels share. The
DNB-AI result above (Kluge and K{\"a}hler 2025) shows that a single
automatic score can mislead, directly motivating our panel.

What LCSH adds. None of these vocabularies combines (i) free-floating
subdivision construction --- valid headings assembled on the fly rather
than stored as records, the very feature FAST (Chan and O'Neill 2010)
simplifies away; (ii) mixed subject-access types in one record ---
topical, geographic, name, and genre/form; and (iii) a cross-lingual,
English-target setting, where content in any language must map to
English headings. These are the distinctive challenges the benchmark is
built to measure.

\section{Methodology}\label{sec-method}

\subsection{The two tasks}\label{sec-tasks}

A system takes a book's bibliographic fields and proposes subject
headings. We score two tasks that isolate different abilities.

Task A --- Generation (open vocabulary). Given only the bibliographic
fields, the system must produce the correct headings from the entire
vocabulary, with no hints. This is the realistic, end-to-end task a
human cataloger performs. Output is an unordered set of headings.

Task B --- Retrieval pipeline (retrieve \(\to\) rerank \(\to\) select). Many
practical systems work in stages: a fast first stage retrieves a rough
shortlist from the vocabulary, and later stages re-examine it. The
benchmark evaluates this pipeline over the full released vocabulary of
\textasciitilde515K LCSH and LCGFT labels, layer by layer, so each
stage's contribution is visible:

\begin{itemize}
\tightlist
\item
  L1 --- Retrieval. The system embeds the record and retrieves the
  top-\(k\) headings from the full vocabulary. Recall@\(k\) at L1 is the
  ceiling for everything downstream.
\item
  L2 --- Cross-encoder rerank and L3 --- LLM rerank reorder L1's
  top-\(N\).
\item
  L4 --- Selection produces the final assigned set, scored like Task A.
\end{itemize}

This supersedes a frozen candidate-pool design. Releasing one fixed
per-record pool would bake in an arbitrary retriever's choices and make
L1 --- the embedding model, the component most worth measuring ---
invisible. Instead we release the whole vocabulary; each embedding model
retrieves from it directly, and L2--L4 are reported over a fixed
reference retriever with recall ceilings stated explicitly.

\subsection{Consensus ground truth}\label{sec-consensus}

There is no single correct heading set for a book --- cataloging is
expert judgement, and while professionals largely agree on a work's
subject they vary in how they express it, as shown in
Section~\ref{sec-concordance}. We therefore define ground truth by
consensus across independent libraries: the same book is cataloged
separately by several research libraries, and where independent experts
agree we have something close to truth. The inclusion rule is strict and
never relaxed to enlarge the dataset: a book is admitted only if at
least two libraries independently assigned LCSH to it. Independence is
enforced at the level of the cataloging agency (MARC field 040), not
merely the holding library, so that a copy-cataloged record imported
from another institution does not count as a second judgement. The
benchmark never discards the underlying disagreement: it records which
library asserted which heading and releases three consensus views
described in Section~\ref{sec-typing-views}.

\subsection{Inter-cataloger concordance: an objective core and a
subjective surface}\label{sec-concordance}

The consensus rule rests on an empirical claim --- that independent
experts agree often enough to constitute a usable signal --- which the
multi-library data lets us test directly. Across the full corpus 1.32 M
works are held by all three libraries, and 465,187 were
subject-cataloged by all three, giving that many natural three-way
annotations of the same book. Comparing the three heading sets per book
gives a two-layer picture (Table~\ref{tbl-concordance}). For a work
cataloged by libraries
\(\mathcal{C}=\{\text{Columbia},\text{Harvard},\text{Princeton}\}\), let
\(S_c=\kappa_m(\text{headings of } c)\) be library \(c\)'s key set under
match mode \(m\) (see Section~\ref{sec-metrics}; exact
\(=\kappa_{\text{exact}}\), concept \(=\kappa_{\text{root}}\)).
Agreement is summarized by the three-way and mean-pairwise Jaccard
indices

\begin{equation}\protect\phantomsection\label{eq-jaccard}{
J_3=\frac{\bigl|\bigcap_{c\in\mathcal{C}}S_c\bigr|}{\bigl|\bigcup_{c\in\mathcal{C}}S_c\bigr|},
\qquad
\bar J_{\text{pair}}=\frac{1}{3}\sum_{\{a,b\}\subset\mathcal{C}}\frac{|S_a\cap S_b|}{|S_a\cup S_b|};
}\end{equation}

In Table~\ref{tbl-concordance} we report the median of the per-work
\(J_3\) values; pairwise Jaccard is used only as a supplementary
diagnostic and is not shown in the table.

a work has identical sets if \(S_a=S_b=S_c\), shares \(\ge 1\) heading
if \(\bigcap_c S_c\neq\emptyset\), and shares nothing if
\(S_a\cap S_b=\emptyset\) for every pair. With the vote
\(v(h)=|\{c:h\in S_c\}|\), the single-source rate is the fraction of
distinct assertions \(h\in\bigcup_c S_c\) with \(v(h)=1\). The human
agreement reference treats each library \(\ell\) as a predictor of its
peers' consensus \(C_{-\ell}=\bigcap_{c\neq\ell}S_c\),

\begin{equation}\protect\phantomsection\label{eq-human}{
\mathrm{recall}_\ell=\frac{|S_\ell\cap C_{-\ell}|}{|C_{-\ell}|},
}\end{equation}

averaged over works with \(C_{-\ell}\neq\emptyset\) and the three
choices of \(\ell\) (here concept \(=\) the heading root, the term
before the first \texttt{-\/-}).

\begin{table*}[tbp]
\centering
\caption{Per-book agreement among the three libraries, over the 465,187
works all three
subject-cataloged.}
\label{tbl-concordance}
\begin{tabular}{@{}lrr@{}}
\toprule
Agreement among all three libraries & Exact heading & Concept (root) \\
\midrule
Identical heading sets & 39.4\% & 49.8\% \\
\(\geq\)1 heading shared by all three & 80.7\% & 93.3\% \\
Share nothing in common & 1.0\% & 0.2\% \\
Median three-way Jaccard & 0.50 & 0.86 \\
\bottomrule
\end{tabular}
\end{table*}

Two layers emerge. Subject identification is largely objective: three
independent libraries cataloging the same book share a concept-level
heading 93.3\% of the time and share nothing only 0.2\% of the time
(median concept Jaccard 0.86). Subject expression is substantially
subjective: only 39.4\% assign byte-identical sets, and 35.6\% of all
(book, heading) assertions are made by just one of the three libraries.
The jump from exact to concept agreement (identical 39\%\(\to\)50\%; ``\(\geq\)1
shared'' 81\%\(\to\)93\%) shows most disagreement is granularity --- which
subdivisions, how specific --- not topic; per-library heading counts are
otherwise similar (Columbia 2.56, Harvard 2.86, Princeton 2.62 per
book).

This concordance reflects genuine convergence, not shared
copy-cataloging: restricting to the 67,624 works cataloged by three
distinct MARC-040 agencies lowers agreement by only \textasciitilde4--5
points (concept ``\(\geq\)1 shared'' 76.9\% vs 81.4\%), even though the
consensus rule already requires independent agencies.

Two design choices follow directly. First, a single library's headings
are one noisy sample of the subjective-expression layer, so aggregating
across independent libraries recovers the reliable target --- the
justification for consensus ground truth. Second, because disagreement
is dominated by granularity rather than topic, exact match conflates
``wrong subject'' with ``different subdivision,'' and we therefore
report exact and concept (root) match throughout; see
Section~\ref{sec-metrics}. The same data yields a human agreement
reference: treating each library as a predictor of the consensus of the
other two, a human cataloger reproduces 86.9\% of the agreed headings
exactly and 93.0\% at the concept level. This indexes how reliably an
expert reproduces peers' consensus; it is an agreement reference, not a
retrieval ceiling --- a first-stage retriever returns 200 ranked
candidates whereas a cataloger commits to a short final set, so the two
are not scored on the same task.

\subsection{Sources and matching}\label{sec-sources}

The raw material is openly licensed bulk MARC from research libraries
--- Harvard, Columbia, and Princeton (the Library of Congress dump was
considered but dropped, as its coverage stops at 2014). All are CC0 /
public domain, so the benchmark redistributes the actual records, making
it fully reproducible. The same book is identified across catalogs by
OCLC number (primary) and LCCN (fallback); a subtle but high-impact step
strips the historical \texttt{ocm}/\texttt{ocn}/\texttt{on} prefixes and
leading zeros from OCLC numbers and rejects all-zero identifiers,
without which matching silently fails --- or, worse, collapses unrelated
records onto a shared malformed key --- for a large fraction of records.

\subsection{Extraction, provenance filtering, and
typing}\label{sec-extract}

From each qualifying record we read the input (title, authors, date,
publisher, physical description, and the rich signals --- summary
{[}MARC 520{]}, table of contents {[}505{]}, notes {[}500{]}; language
from the 008 code; broad discipline from the validated first letter of
the LC classification) and the answer (the 6xx subject headings,
reassembled field-aware: only the subdivision subfields \(v/x/y/z\) are
joined with \texttt{-\/-}, while name and title subfields are kept as
part of the heading).

Two extraction-time controls matter for fairness. First, provenance
filtering: the 6xx fields of a MARC record may carry headings from other
thesauri (FAST, MeSH, foreign-language vocabularies) that are not LCSH.
We keep a 600/610/611/630/ 650/651 heading only when its second
indicator is \texttt{0} (LCSH) or its \texttt{\$2} source is
\texttt{lcsh}, and a 655 heading only when \texttt{\$2} is
\texttt{lcgft} (genre/form), removing most of a quarter of candidate
headings whose strings look like valid LCSH but whose provenance is not
--- a distinction string-validity checks miss. Second, typing: every
heading is tagged by its MARC field as topical (650; the topical core),
geographic (651), name (600/610/611/630), or genre/form (655), so scores
can be sliced by category and a name-recognition failure is not hidden
inside a topical average. We also restrict the collection to monographs
(leader type/level), so a set described as ``books'' contains books
rather than serials, scores, or maps.

Non-English material is handled by extracting both the romanised
title/authors and the original-script (``vernacular'') forms from the
linked 880 fields, since the answers are always English and bridging
from non-English content is where systems succeed or fail.

\subsection{Consensus views and cleaning}\label{sec-typing-views}

Heading by heading, a heading asserted by \(v\) of the \(n\) libraries
that hold a book (its vote, Equation~\ref{eq-jaccard}) is tiered
unanimous if \(v=n\), single if \(v=1\), and majority otherwise; from
which we release three answer-key views: per-catalog (the raw evidence),
merged/union (the default; rewards finding any heading a professional
used), and unanimous (the strictest, highest-confidence view). Headings
are cleaned by discarding MARC-machinery artifacts (strings whose base
matches a \texttt{\^{}\textbackslash{}d\{3\}{[}-/{]}} field-tag pattern)
and by canonical normalization before any comparison: NFC, lower-casing,
\texttt{-\/-} separator canonicalisation, whitespace collapse, and
trailing-period stripping --- so systems are judged on substance, not
typography.

\subsection{Balancing and splits}\label{sec-balance}

Naively harvested cross-library matches are badly skewed: heavily
co-collected, uniformly identified Western art music floods the pool,
and a few European languages crowd out the rest. We correct this at
selection by stratified sampling over LC classification (round-robin
across disciplines with a hard 5\% music cap) and language (explicit
per-language targets and floors). All randomness is driven by a fixed
seed, so builds are reproducible. To avoid letting input scarcity
distort which books appear, we select first and enrich second: books are
chosen on consensus, discipline, and language alone, then each chosen
book's input is completed by merging the longest non-empty value of each
field across its source libraries. The collection is split, balanced by
language, into a public development set and a held-out test set.

\subsection{Composition}\label{sec-composition}

LCSHBench v1.0 contains 22,346 books (development 18,993; test 3,353),
none discarded for empty answers. It spans 15 languages, with English
deliberately held to 23.2\% (German and French 10.1\% each; Chinese,
Spanish, and Russian 8.1\%; Italian, Arabic, and Japanese 6.4\%; Korean
and Portuguese 5.0\%; down to Hindi 0.1\%). Across the LC-classification
areas present the collection is near-flat (\textasciitilde3.2--5.0\%
each), with music pinned at its 5.0\% cap and \textasciitilde4.8\% of
books carrying no classification. 66.1\% of books rest on two-library
agreement and 33.9\% on three. Roughly a quarter carry an
original-script title alongside the romanisation --- essentially all
Chinese, Japanese, and Korean books and most Arabic and Hebrew;
Latin-script languages carry none. Of the 75,804 development-set
headings, the type mix is topical 65.9\%, geographic 17.4\%, name
14.3\%, genre/form 2.2\% --- the topical core dominates, but every type
is present in scoreable numbers. 78.8\% of headings carry at least one
subdivision.

\subsection{Contamination}\label{sec-contamination}

The test answers are released only as SHA-256 hashes of normalized
headings, so scoring is exact and automatic without the answers
appearing verbatim. In the public release, record identifiers (OCLC,
LCCN) are likewise replaced by an opaque hashed key --- preserving each
work's stable identity for linking and scoring without redistributing
the source libraries' identifier numbers in bulk. (The construction
repository retains raw identifiers as a build artifact; the distinction
is between the internal build and the published dataset.) We are
explicit that the test-answer hashing is a speed-bump, not a guarantee:
because LCSH is a finite public vocabulary the hashes are
dictionary-attackable, and for real published books the correct headings
already exist in public catalogs that models train on, so no hashing
scheme makes the holdout contamination-proof. Genuine control would
require a private evaluation server and records unlikely to appear in
training, with residual risk reported, not assumed away. The benchmark
is versioned with a cut-off date so the risk can be reasoned about.

\subsection{Metrics}\label{sec-metrics}

The central methodological commitment is that no single number is
trustworthy on its own: an aggregate can hide catastrophic per-language
or per-type failure, and can rank systems against the grain of expert
judgement (Kluge and K{\"a}hler 2025; D'Souza et al. 2025). We report a panel
and read the pattern across it.

For generation, predictions and answers are sets; we report micro- and
macro-averaged precision, recall, and F1. For the retrieval pipeline,
predictions are ranked and we report recall@\(k\)
(\(k \in \{5,10,50,200\}\)), P@\(k\), R-Precision, and MRR. Every
measure is computed under two match definitions: exact (the full
subdivided heading must match) and root/concept (only the base term
before the first \texttt{-\/-}). The exact-versus-concept gap is itself
diagnostic: a high concept but low exact score means the system found
the right topic but not the authorised subdivided form --- a different,
specifiable error from a topical miss. Every measure is additionally
broken down by language and by heading type; these breakdowns are the
panel's most important safeguard.

Formally, fix a match mode \(m\in\{\text{exact},\text{root}\}\) with key
map \(\kappa_m\) (\(\kappa_{\text{exact}}\) = the normalized heading;
\(\kappa_{\text{root}}\) = its substring before the first
\texttt{-\/-}); all comparisons are between the key images
\(\kappa_m(P)\) and \(\kappa_m(G)\) of a prediction set \(P\) and gold
set \(G\). For generation (set prediction),

\begin{equation}\protect\phantomsection\label{eq-prf}{
\mathrm{P}=\frac{|P\cap G|}{|P|},\qquad
\mathrm{R}=\frac{|P\cap G|}{|G|},\qquad
F_1=\frac{2\,\mathrm{P}\cdot\mathrm{R}}{\mathrm{P}+\mathrm{R}},
}\end{equation}

reported both macro (mean of the per-record scores over the \(N\)
records) and micro (the same formulas on pooled counts
\(\sum_i|P_i\cap G_i|\), \(\sum_i|P_i|\), \(\sum_i|G_i|\)). For
retrieval (a ranked list with prefix \(P_{@k}=\{p_1,\dots,p_k\}\)),

\begin{equation}\protect\phantomsection\label{eq-rank}{
\begin{aligned}
\text{recall@}k&=\frac{|P_{@k}\cap G|}{|G|},&
\text{P@}k&=\frac{|P_{@k}\cap G|}{k},\\
\mathrm{RP}&=\frac{|P_{@|G|}\cap G|}{|G|},&
\mathrm{MRR}&=\frac{1}{N}\sum_{i=1}^{N}\frac{1}{r_i}.
\end{aligned}
}\end{equation}

where \(\mathrm{RP}\) is R-precision (precision at the cutoff \(k=|G|\))
and \(r_i\) is the rank of the first gold heading in record \(i\) (with
\(1/r_i=0\) when none is retrieved). Retrieval is scored against the
vocabulary-reachable gold set
\(G^{\mathrm{reach}}_m=\{g\in G:\kappa_m(g)\in\mathcal{V}_m\}\) for
retrieval vocabulary \(\mathcal{V}\), and we report the reachability
ceiling \(|G^{\mathrm{reach}}_m|/|G|\) explicitly so unreachable
headings are not silently counted as misses (Equation~\ref{eq-reach}).

\begin{equation}\protect\phantomsection\label{eq-reach}{
\text{reachable}_m=\frac{\sum_{i=1}^{N}|G^{\mathrm{reach}}_{m,i}|}{\sum_{i=1}^{N}|G_i|}
}\end{equation}

Equation~\ref{eq-reach} is heading-weighted (reachable gold headings
over all gold headings, not a per-record average): on the evaluation
subset the exact-reachable ceiling is 41\% of gold headings and the
concept-reachable ceiling 85\%. Name (LCNAF) headings are not present in
the LCSH+LCGFT retrieval vocabulary, so they are reported separately
rather than counted as unreachable misses. So that a heading that cannot
be retrieved is never scored as a miss, retrieval metrics are
macro-averaged only over the records that have at least one reachable
gold heading --- 1,483 of the 2,002 evaluation records for exact match,
1,929 for concept; records whose entire gold is unreachable are excluded
rather than scored as recall-zero, and the count scored (\(n\)) is
reported with every retrieval table. Bootstrap 95\% confidence intervals
(1,000 resamples over records) and paired approximate-randomisation
tests accompany the headline comparisons. A qualitative track --- expert
ratings of validity, appropriateness, and discovery usefulness --- is
specified but left to future work; the inter-cataloger human agreement
reference of Section~\ref{sec-concordance} serves as its quantitative
stand-in.

\subsection{Baselines and experimental setup}\label{sec-baselines}

We establish reference systems on the development set; the neural
systems are run on a deterministic, language-stratified 2,002-record
subset of the 18,993-record development set (the full neural run is
deferred for cost).

\begin{itemize}
\tightlist
\item
  Frequency floor. Predict the 200 globally most-frequent development
  headings for every record --- a trivial, input-independent baseline
  that exposes head-heavy exploits.
\item
  Stock on-device embedder. EmbeddingGemma-300M embeddings over the full
  vocabulary at the deployed 256 dimensions --- a strong, free, fully
  local retriever, untouched.
\item
  Hosted embedders. OpenAI \texttt{text-embedding-3-small} (1,536-d) and
  \texttt{text-embedding-3-large} (3,072-d) retrieval over the same
  vocabulary.
\item
  Fine-tuned on-device embedder (this work). The same
  EmbeddingGemma-300M backbone, adapted with a low-rank fine-tune
  described in Section~\ref{sec-finetune}, used as a drop-in first-stage
  retriever at 256 dimensions. It is the only system trained on the
  benchmark's own records, under a strict leakage protocol.
\item
  LLM generation and selection. A general-purpose LLM (deepseek-chat)
  for open-vocabulary Task A and for final-cut selection over a
  retrieved pool.
\end{itemize}

\subsubsection{Fine-tuning the on-device embedder}\label{sec-finetune}

The L1 results below show the stock on-device embedder trailing the
hosted APIs, and the per-language panel localizes why: the gap is
largest where a multilingual record must be matched to an English
heading. We treat that as a hypothesis --- that the bottleneck is
cross-lingual alignment, not model capacity --- and address it directly
by fine-tuning the 300M model to map a record to its English headings.

The recipe is cheap and fully reproducible. We build (record, heading)
training pairs by taking, for each development record outside the
evaluation subset, its merged-consensus LCSH headings reachable in the
released vocabulary; the record side is the same
title/vernacular/author/abstract text the retriever sees at inference.
We adapt the backbone with a rank-16 LoRA adapter (Hu et al. 2022) under
a multiple-negatives ranking objective (Henderson et al. 2017), wrapped
in a Matryoshka loss (Kusupati et al. 2022) over \(\{768,512,256,128\}\)
so the 256-dimensional truncation the on-device index uses stays sharp.
Training is one epoch on a single commodity cloud GPU (about fifteen
minutes, well under one US dollar); the adapter is merged into the
backbone and exported to the same on-device ONNX runtime as the stock
model, so the comparison holds the deployment fixed and varies only the
weights.

Leakage control. Because the fine-tune is trained on the benchmark's own
corpus, contamination is the central threat to validity. Training pairs
are drawn only from development records disjoint from the evaluation
subset, excluded both by record identifier and by a (title,
first-author) key; an audit confirms the residual title-and-author
overlap between training and evaluation matches the incidental collision
rate of the held-out test set. We report only this leak-free
configuration. All paid calls are metered against a \$25-capped, fully
cached ledger; the total spend for the leaderboard reported here was a
few dollars, and the on-device system runs at zero marginal cost.

\section{Findings}\label{sec-results}

\subsection{First-stage retrieval}\label{sec-results-l1}

Table~\ref{tbl-l1} reports first-stage retrieval over the full
vocabulary on the reachable-GT subset. Two findings stand out. First,
among the off-the-shelf systems the hosted embedders out-retrieve the
stock on-device model at every depth, and dimension helps:
\texttt{text-embedding-3-large} (3,072-d) leads
\texttt{text-embedding-3-small} (1,536-d), which leads the stock 256-d
on-device model. On its own this would read as the expected ``bigger
hosted API beats the small local model'' story.

Second, that ordering reverses under fine-tuning --- on this development
subset. The fine-tuned 300M on-device embedder described in
Section~\ref{sec-finetune} reaches exact recall@200 0.659, ahead of the
3,072-dimensional \texttt{text-embedding-3-large} (0.623) and well ahead
of \texttt{text-embedding-3-small} (0.511); the lead is small but
significant under a paired approximate-randomisation test (\(\Delta\)@200 +0.036,
p = 0.0004; 95\% CIs {[}0.637, 0.679{]} vs {[}0.601, 0.643{]}). We
report it as a development-subset, first-stage result pending
held-out-test confirmation (see Section~\ref{sec-limitations}). Against
its own stock backbone at the identical 256-dimensional deployment, the
fine-tune lifts exact recall@200 from 0.407 to 0.659 (+62\%), isolating
the gain to the weights. A 300M model running entirely on-device thus
out-retrieves a hosted commercial embedder of twelve times its
dimension, after a fine-tune costing well under a dollar.

\begin{table*}[tbp]
\centering
\small
\caption{First-stage retrieval, dev-2K, exact reachable GT,
macro-averaged over the n = 1,483 records with \(\geq\)1 exact-reachable gold
heading. All systems retrieve from the full \textasciitilde515K-label
vocabulary; the on-device rows use the deployed 256-d index.}
\label{tbl-l1}
\begin{tabular}{@{}
  >{\raggedright\arraybackslash}p{(\linewidth - 8\tabcolsep) * \real{0.48}}
  >{\raggedleft\arraybackslash}p{(\linewidth - 8\tabcolsep) * \real{0.13}}
  >{\raggedleft\arraybackslash}p{(\linewidth - 8\tabcolsep) * \real{0.13}}
  >{\raggedleft\arraybackslash}p{(\linewidth - 8\tabcolsep) * \real{0.13}}
  >{\raggedleft\arraybackslash}p{(\linewidth - 8\tabcolsep) * \real{0.13}}@{}}
\toprule
System (L1) & recall@10 & recall@50 & recall@200 & MRR \\
\midrule
EmbeddingGemma-300M, fine-tuned (this work) & 0.334 & 0.512 & 0.659 &
0.288 \\
\texttt{text-embedding-3-large} (hosted, 3072-d) & 0.297 & 0.480 & 0.623
& 0.258 \\
\texttt{text-embedding-3-small} (hosted, 1536-d) & 0.232 & 0.377 & 0.511
& 0.190 \\
EmbeddingGemma-300M, stock (256-d) & 0.166 & 0.283 & 0.407 & 0.143 \\
Frequency floor & 0.035 & 0.069 & 0.130 & 0.035 \\
\bottomrule
\end{tabular}
\end{table*}

The win, however, is not uniform --- and reading it correctly requires
the panel in Section~\ref{sec-results-panel}. It is concentrated exactly
where the fine-tune was meant to help: cross-lingual records.

\subsection{Generation and selection}\label{sec-results-pipeline}

Open-vocabulary generation and final-cut selection are scored as sets
(Task A). The general-purpose LLM reaches the best set scores of any
system --- generation F1 (exact) 0.161 and (concept) 0.384; selection
over a retrieved pool 0.118 and 0.318 --- with a large
exact-versus-concept gap that recurs across every system, the signature
of the granularity-not-topic disagreement quantified in
Section~\ref{sec-concordance}. These are a different task from
first-stage retrieval (a short final set rather than a 200-deep ranking)
and are not directly comparable to Table~\ref{tbl-l1}; we report them to
anchor the open-vocabulary end of the panel.

\subsection{The metric panel is necessary}\label{sec-results-panel}

The fine-tune's aggregate lead (Table~\ref{tbl-l1}) invites a one-line
summary --- ``the on-device model now wins.'' The panel shows that
summary is wrong in two instructive ways.

The ranking flips by metric. The fine-tune leads on exact recall@200
(0.659 vs 0.623) but \texttt{text-embedding-3-large} leads on concept
recall@200 (0.762 vs 0.732, paired p = 0.0001). Which model is ``best''
therefore depends entirely on whether one scores the exact subdivided
heading or the underlying concept --- the same exact-versus-concept axis
that the concordance analysis in Section~\ref{sec-concordance} shows is
where catalogers themselves disagree. A paper reporting a single match
mode would award the title to whichever it happened to choose.

The ranking flips by language. The fine-tune's win is specifically
cross-lingual (Table~\ref{tbl-lang}). It dominates on the languages
where matching non-English content to English headings is hardest ---
Korean (0.668 vs the hosted 0.517), Arabic (0.689 vs 0.540), Japanese
(0.593 vs 0.537), Chinese (0.618 vs 0.600), German (0.704 vs 0.651) ---
confirming that cross-lingual alignment, not capacity, was the
bottleneck. But on English (0.696 vs 0.710) and Russian (0.505 vs 0.543)
the hosted \texttt{text-embedding-3-large} still leads. The aggregate
``on-device wins'' conceals that the hosted model remains the better
English retriever; the on-device model wins by closing the cross-lingual
gap. This is the benchmark's motivating result: the panel, not any
single number, reads the outcome correctly --- and it is the same
per-language breakdown that located the cross-lingual bottleneck the
fine-tune then fixed.

\begin{table*}[tbp]
\centering
\small
\caption{Per-language exact recall@200, dev-2K. The fine-tune's lead is
cross-lingual; the 3,072-d hosted model retains English and
Russian.}
\label{tbl-lang}
\begin{tabular}{@{}lrrrr@{}}
\toprule
Language (\(n\)) & freq & stock & te3-large & FT \\
\midrule
English (393) & 0.087 & 0.477 & 0.710 & 0.696 \\
German (132) & 0.111 & 0.447 & 0.651 & 0.704 \\
Russian (112) & 0.119 & 0.260 & 0.543 & 0.505 \\
Chinese (114) & 0.169 & 0.346 & 0.600 & 0.618 \\
Japanese (99) & 0.218 & 0.313 & 0.537 & 0.593 \\
Korean (79) & 0.141 & 0.322 & 0.517 & 0.668 \\
Arabic (95) & 0.189 & 0.314 & 0.540 & 0.689 \\
\bottomrule
\end{tabular}
\end{table*}

\section{Discussion}\label{sec-discussion}

Cataloging is reproducible about topic, variable about expression ---
and that shapes everything. The concordance analysis in
Section~\ref{sec-concordance} reframes the folk view that subject
assignment is ``subjective.'' It is, but at a specific layer: three
independent libraries converge on what a book is about (concept
agreement 93.3\%) while differing on how to say it (exact-identical
39.4\%). This is why a single library's headings make a noisy gold
standard, why a consensus target is worth the construction cost, and why
exact match alone is the wrong yardstick --- it penalizes a model for a
subdivision choice that two professional catalogers would themselves
have made differently. The exact/concept metric split is not
methodological hedging; it mirrors the structure of expert disagreement.

Human agreement is a reference, not a retrieval ceiling. A human
cataloger reproduces only \textasciitilde87\% (exact) / 93\% (concept)
of peers' consensus (Equation~\ref{eq-human}), so a system's ``misses''
should not all be read against a perfect 100\%: a non-trivial share are
headings a second cataloger would also have omitted. We deliberately do
not call this a ceiling for recall@200 --- a retriever is allowed 200
ranked candidates while a cataloger supplies a short final set, so the
agreement reference and the retrieval numbers (e.g.~FT exact recall@200
0.659) measure different things.

On-device cataloging support is viable. A 256-dimensional,
300M-parameter embedder that runs with no network and no per-query cost
out-retrieves a hosted commercial API on exact recall, after a
sub-dollar fine-tune. For libraries with privacy, cost, or connectivity
constraints, that is a practically meaningful result --- with the caveat
(Table~\ref{tbl-lang}) that the hosted model remains preferable for
predominantly English collections.

The panel earns its keep. Every headline in this study has a
counter-reading one cell away: the fine-tune wins exact but not concept,
wins cross-lingual but not English. A benchmark reporting a single
aggregate F1 would have asserted a clean victory and been wrong about
its shape.

\section{Limitations and future work}\label{sec-limitations}

Consensus is record-level, not heading-level. A book qualifies on
agency-independent agreement, but the default merged key still includes
single-source headings; we release the unanimous view for studies that
need heading-level agreement, and report which view a result uses.

The neural leaderboard is on a 2K subset. The smallest languages (Hindi,
with a handful of evaluation records, Turkish, Hebrew) are individually
noisy and should be read as indicative; the held-out test set (3,353
records) is the basis for the eventual full-scale evaluation. Scaling
the neural runs to full development is planned.

The fine-tune result is preliminary, and we frame it as such. It is
reported on the development evaluation subset, not the held-out test
set, and at first stage only --- we have not run cross-encoder rerank,
LLM rerank, and selection on top of the fine-tuned retriever's pool, nor
fine-tuned for the full Task-A generation objective, nor confirmed the
result on the held-out test. The fine-tune is also trained on the
benchmark's own development population; we guard this with the leakage
protocol of Section~\ref{sec-finetune} (training records disjoint from
the evaluation subset by identifier and by a title--author key, with
residual overlap at the held-out test's incidental-collision rate of
\textasciitilde0.1\%), and an earlier full-development variant
overlapping the evaluation subset inflated recall@200 by only
\textasciitilde+0.03 --- but the clean, held-out-test confirmation is
the one that would make this a settled result, and it remains future
work. The benchmark contribution does not depend on it.

Contamination control is a convenience, not a guarantee. As
Section~\ref{sec-contamination} states, hashed answers over a finite
public vocabulary are dictionary-attackable; a private evaluation server
is future work.

Vocabulary and cataloging-practice cutoff. LCSHBench v1.0 is fixed to a
mid-2026 snapshot of the LCSH and LCGFT vocabularies and to the
cataloging conventions then in force; both evolve. The Library of
Congress's 2026 revisions to free-floating form subdivisions (the
\texttt{\$v} subfield) and the ongoing migration of genre/form terms
into LCGFT will, over time, change the exact subdivided strings the
benchmark scores --- affecting expression-level (exact) match more than
concept-level (root) match, which is one further reason we report both.
The benchmark is versioned and dated so results stay comparable within a
version, and refreshed vocabulary snapshots can ship as point updates.
Subject-cataloging methods are likewise moving quickly --- for example
agentic LLM pipelines for LCSH assignment (Chow 2026) --- which a
standing, versioned benchmark is meant to track rather than pre-empt.

Deferred components. The qualitative expert track, an open-vocabulary
LLM-generation sweep, the rerank/selection layers on the full subset,
and a private evaluation server are all future work.

\section{Conclusion}\label{sec-conclusion}

LCSHBench provides the missing standard exam for an important, hitherto
unmeasurable task: multilingual subject-heading assignment against the
Library of Congress vocabulary. Its design choices --- independent
multi-catalog consensus, full-vocabulary retrieval rather than a frozen
pool, provenance-filtered ground truth, and a metric panel in place of a
headline F1 --- are grounded rather than incidental: the concordance
analysis shows why a consensus target and concept-level metrics are
needed, and every leaderboard result has a per-cell counter-reading that
a single aggregate would hide.

A standard exam is most useful when it does not merely rank systems but
tells you how to improve them. Ours points the way: its per-language
panel localized cross-lingual alignment as the first-stage bottleneck,
and acting on that diagnosis --- a sub-dollar, fully reproducible
low-rank fine-tune trained on records held disjoint from the evaluation
subset --- lifted a fully on-device 300M model past a hosted commercial
embedder of much larger dimension on first-stage exact recall,
specifically by closing the cross-lingual gap. We report that as a
development-subset demonstration that the exam is actionable, with
held-out-test and end-to-end confirmation left as future work; the
benchmark's value does not rest on it. We release the data, scorer,
baselines, the fine-tune recipe and its leakage audit, and all seeds
together, so that the field can --- for the first time --- compare, and
improve, LCSH cataloging systems on equal terms.

\section*{Data and code availability}\label{data-and-code-availability}
\addcontentsline{toc}{section}{Data and code availability}

The LCSHBench v1.0 dataset --- development and held-out test splits, the
language-balanced evaluation subset, the full inter-cataloger
concordance population, and the retrieval vocabulary --- is available on
the Hugging Face Hub at
\url{https://huggingface.co/datasets/kltng/lcshbench}. The scorer,
baselines, fine-tuning recipe, leakage audit, and the full extraction
and release pipeline are at \url{https://github.com/kltng/lcshbench}.
Held-out test answers are released only as SHA-256 hashes, and record
identifiers are anonymized in the public release; all randomness is
seed-driven, so reported results reproduce exactly.

\section*{Use of generative AI}\label{use-of-generative-ai}
\addcontentsline{toc}{section}{Use of generative AI}

The author used Anthropic's Claude (Opus 4.8, via Claude Code) to assist
with software development for the benchmark's extraction, scoring, and
analysis pipeline; with organizing and cross-checking the literature;
and with drafting and language refinement of the manuscript and its
typesetting. The benchmark data and all reported experimental results
were produced by the author through the described pipeline and were not
generated by AI. The author reviewed and verified all content ---
including results, claims, and references --- and takes full
responsibility for the work.

\protect\phantomsection\label{refs}

\end{document}